\documentclass[onecolumn,showpacs,preprintnumbers,amsmath,amssymb]{revtex4}

\usepackage{graphicx}
\usepackage{dcolumn}
\usepackage{bm}

\begin{document}

\preprint{APS/123-QED}

\title{Light deflection in the second post-Newtonian approximation \\of
scalar-tensor theory of gravity}

\author{Peng Dong}%
\email{dongpeng@pmo.ac.cn}%
\affiliation{Center for Gravitation and Cosmology, Purple Mountain
observatory,
Chinese Academy of Sciences, Nanjing, 210008 China.}%
\author{Wei-Tou Ni}%
\email{wtni@pmo.ac.cn}%
\affiliation{Center for Gravitation and Cosmology, Purple Mountain
observatory, Chinese Academy of Sciences, Nanjing, 210008 China.}%
\affiliation{National Astronomical Observatories, Chinese Academy of
Sciences,
 Beijing, 100012 China.
}%

\date{\today}

\begin{abstract}
In this paper, we use the metric coefficients and the equation of
motion in the 2nd post-Newtonian approximation in scalar-tensor
theory including intermediate range gravity to derive the deflection
of light and compare it with previous works. These results will be
useful for precision astrometry missions like GAIA (Global
Astrometric Interferometer of Astrophysics), SIMS (The Space
Interferometry Mission) and LATOR (Laser Astrometric Test Of
Relativity) which aim at astrometry with microarcsecond and
nanoarcsecond accuracies and need 2nd post-Newtonian framework and
ephemeris to determine the stellar and spacecraft positions.
\end{abstract}

\pacs{Valid PACS appear here}
\maketitle

\section{INTRODUCTION}

The relativistic light deflection passing near the solar rim is 1.75
as (arcsec). The first post-Newtonian approximation is valid to
$10^{-6}$ and the second post-Newtonian is valid to $10^{-12}$ of
relativistic effects such as light deflection in the solar system.
For astrometry mission to measure angles with accuracy in the nas to
$\mu$as range, 2nd post-Newtonian approximation of relevant theories
of gravity is required both for the angular measurement and for
relativistic gravity test. The scalar-tensor theory is widely
discussed and used in tests of relativistic gravity. In order to
confront the predictions of scalar-tensor theory with experiment in
the solar system, it is necessary to compute it's second
post-Newtonian approximation and certain gravitational effects such
as deflection of light, time delay of light and perihelion shift in
this approximation. The second post-Newtonian contribution for light
ray has been discussed for a long time by  and
\cite{EpsteinShapiro80}, \cite{RichterMatzner82} and by others
later. In this paper, we use the metric coefficients we obtained
earlier (\cite{Yi Xie07}) to compute the deflection in the second
post-Newtonian approximation considering the velocity of the
observer (spacecraft).\\

\section{Metric coefficients}

The calculation of light deflection to 2PN approximation requires
knowledge of  terms in the metric to order $(v/c)^{4}$. For the
scalar-tensor theory, the metric coefficients are
\begin{eqnarray}
g_{00}=1-2U+2(1+\bar{\beta})U^{2},\quad g_{0i}=L_{i},\quad
g_{ij}=-\delta_{ij}[1+2(1+\bar{\gamma}) U+\frac{3}{2}(1+\Lambda)
U^{2}]
\end{eqnarray}
in the global coordinates. Note that $U$ is given by
\begin{eqnarray}
U=\int\frac{\rho(\vec{x}',t)}{|\vec{x}-\vec{x}'|}d^{3}x'
-\frac{\bar{\gamma}\xi_{1}}{4\pi}\int\frac{d^{3}x'}{|\vec{x}-\vec{x}'|}\int\frac{\rho(\vec{x}'',t)}{|\vec{x}'-\vec{x}''|}
\exp[-\xi_{1}(\vec{x}'-\vec{x}'')]d^{3}x''
\end{eqnarray}
and the parameters $\bar{\gamma}$, $\bar{\beta}$, $\xi_{1}$ and
$\Lambda$ are given in \cite{Yi Xie07} by
\begin{eqnarray}
\bar{\gamma}=-\frac{1}{\omega_{0}+2},\quad
\bar{\beta}=\frac{\omega_{1}}{(2\omega_{0}+3)(2\omega_{0}+4)^{2}},\quad
\Lambda=\frac{15}{6}\bar{\gamma}+\frac{4}{3}(\bar{\gamma}^{2}+\bar{\beta}),\quad
\xi_{1}=-4\frac{\bar{\gamma}\lambda_{2}\phi_{0}^{2}}{2+\bar{\gamma}}
\end{eqnarray}

\section{Deflection angle}

The basic equations of light ray read
\begin{eqnarray}
g_{\mu\nu}k^{\mu}k^{\nu}=0,\quad
dk^{\mu}/dt+\Gamma^{\mu}_{\rho\sigma}k^{\rho}k^{\sigma}=0
\end{eqnarray}
where $k^{\mu}\equiv dx^{\mu}/dt$. Consider a light signal emitted
at $(\vec{x}_{0},t_{0})$ in an initial direction described by the
unit vector $\hat{n}$ satisfying $\hat{n}\cdot\hat{n}=1$ and let its
have the form
\begin{eqnarray}
\vec{x}(t)=\vec{x}_{0}+\hat{n}(t-t_{0})+\vec{x}_{p}(t)+\vec{x}_{pp}(t)
\end{eqnarray}
where $\vec{x}_{p}(t)$and $\vec{x}_{pp}(t)$ are the first and second
post-Newtonian correction respectively. For the rotating sun, we
obtain the solution needed to the second-order approximation by the
iterative method.

Consider an observer (satellite, spacecraft) with the four-velocity
$u^{\mu}$ who receives the signals from two different sources. The
angle $\alpha$ between the directions of two incoming photons is
given by the following expression:
\begin{eqnarray}
\cos \alpha=k^{\mu}_{I}P^{\rho}_{\mu}k^{\sigma}_{II}P_{\rho\sigma}
|k^{\mu}_{I}P^{\rho}_{\mu}|^{-1}|k^{\sigma}_{II}P_{\rho\sigma}|^{-1}=f(g_{\mu\nu},u^{\sigma},k^{\rho}_{I},k^{\tau}_{II}),
\end{eqnarray}
where $P_{\mu\nu}=g_{\mu\nu}+u_{\mu}u_{\nu}$ is a projection
operator. Defining the angle $\delta\alpha$ to be the deflected
angle from the original angle $\alpha_{0}$, and expanding $\cos
\alpha$ around $\alpha_{0}$ to the second order, we have
\begin{eqnarray}
\delta\alpha_{p}=\cot \alpha_{0}-f\csc
\alpha_{0},\delta\alpha_{pp}=\cot \alpha_{0}-f\csc
\alpha_{0}-\frac{1}{2}(\delta\alpha_{p})^{2}\cot
\alpha_{0}-\delta\alpha_{p}
\end{eqnarray}
where $\delta\alpha_{p}$ and $\delta\alpha_{pp}$ are the deflection
angles for the first and second post-Newtonian approximations. After
a straightforward but lengthy calculation, we have
\begin{eqnarray}
\delta\alpha=2(2+\bar{\gamma})(M/R)+(2+\bar{\gamma})\bar{\gamma}\xi_{1}(M/2\pi
R)-(2+\bar{\gamma})(MR/2r^{2}_{os})+2(2+\bar{\gamma})J_{2}(M/R)
\nonumber\\\pm 2(2+\bar{\gamma})(J/R^{2})
+[(30+31\bar{\gamma}+8\bar{\gamma}^{2})\pi-16(2+\bar{\gamma})^{2}](M^{2}/8R^{2}),
\end{eqnarray}
for light passing the solar limb in the equatorial plane from
outside the solar system. Here $M$ is the solar mass, $R$ the solar
radius, $J$ the solar angular momentum and $J_{2}$ the solar
quadrupole moment parameter. The second term comes from the
intermediate-range force and other terms agree with the former
works.

The 2PN light trajectory obtained here is useful for obtaining 2PN
range of deep space laser ranging missions ASTROD I and ASTROD
(\cite{Ni04}).

A long paper on this topic will be presented in the future.\\

We thank the National Natural Science Foundation (Grant Nos 10475114
and 10778710) and the Foundation of Minor Planets of purple Mountain
Observatory for support.

\end{document}